\documentclass[aps,floatfix,showpacs,preprintnumbers,amsmath,amssymb]{revtex4}
\usepackage{natbib}
\usepackage{dcolumn}
\usepackage{graphicx}
\usepackage{enumerate}
\usepackage{inputenc}
\usepackage{fontenc}
\usepackage{float}
\usepackage{hyperref}

\newcommand{\mnras}{Mon.~Not.~R.~Astron.~Soc.}

\newcommand{\phrd}{Phys.~Rev.~D.}
\newcommand{\jcap}{J.~Cosmol.~Astropart.~Phys.}
\newcommand{\be}{\begin{equation}}
\newcommand{\ee}{\end{equation}}
\newcommand{\bea}{\begin{eqnarray}}
\newcommand{\eea}{\end{eqnarray}}
\providecommand{\abs}[1]{\lvert#1\rvert}
\def\[{\begin{equation}}
\def\]{\end{equation}}
\begin{document}
\title{Spatial curvature and cosmological tests of general relativity}
\author{Jason N.  Dossett$^1$\footnote{Electronic address: jdossett@utdallas.edu}, Mustapha Ishak$^1$\footnote{Electronic address: mishak@utdallas.edu}}
\affiliation{
$^1$Department of Physics, The University of Texas at Dallas, Richardson, TX 75083, USA;}
\date{\today}
\begin{abstract}
It is well-known that allowing for spatial curvature affects constraints on cosmological parameters such as the dark energy equation of state parameters.  Here we study the effect of curvature on constraints on parameters used to test general relativity (GR) at cosmological scales, commonly known as modified growth (MG) parameters, as while current data taken in the context of the $\Lambda$CDM model points to a universe that is flat or very close to it, this constraint may not hold in modified theories of gravity. Using the latest cosmological data sets we find that MG parameters are correlated with the curvature parameter $\Omega_k$ and the constraints on the MG parameters are weakened compared to when $\Omega_k$ is not included in the parameter analysis. We next use various future simulated data sets, including cosmic microwave background, weak lensing, and Integrated Sachs-Wolfe-galaxy cross-correlations, where the fiducial model is spatially curved but we assume a flat model when fitting the MG parameters.  We find the assumption of a spatially flat model on a spatially curved universe does indeed cause an artificial shift in the constraints on the MG parameters, in some cases even producing an apparent deviation from GR in the MG parameter space.  For our simulated data, tension with GR begins to manifest itself for fiducial models with $\abs{\Omega_k} \geq 0.02$ and apparent deviations appear for $\abs{\Omega_k} \geq 0.05$. We find that for negatively curved models the apparent deviation is more significant.  The manifestation of this apparent deviation from GR due to the assumption of spatial flatness above leads one to conclude that, when using future high-precision data to perform these tests, spatial curvature must be included in the parameter analysis along with the other core cosmological parameters and the MG parameters.
\end{abstract}
\maketitle
\section{introduction}
In an era of precision cosmology several studies have shown that allowing curvature can affect constraints on other cosmological parameters, particularly the equation of state of dark energy (see for example Refs. \cite{r1, r2, r3, r4, r5, r6}).  With the current interest in constraining modified growth (MG) parameters to test general relativity on cosmological scales, one might wonder what effect curvature will have on these constraints and their implications.  

Indeed the effort of understanding the cause of cosmic acceleration by measuring the equation of state of dark energy has  been joined by studies in constraining deviations from general relativity at cosmological scales \cite{r7, r8, r9, r10, r11, r12, r13,r14, r15, r16, r17, r18, r19, r20, r21, r22, r23, r24, r25, r26, r27, r28, r29, r30, r31,r32} and most recently by using current data to constrain the MG parameters \cite{r33, r34, r35,r36, r37,r38, r39, r40,r41} to determine if cosmic acceleration could be due to some extension to general relativity rather than an unknown dark energy component pervading the universe.  So far, all constraints on MG parameters have shown that current observations are consistent with general relativity.  However, these current constraints are not stringent enough to provide any definitive answer. With the exception of Ref. \cite{r36}, to date these studies have been carried out assuming a flat universe. It is worth exploring what impact, if any, allowing curvature has on both constraining the MG parameters and the performance of these tests.   

In this paper, we explore the effect of spatial curvature on testing general relativity (GR) at cosmological scales.  We do this because though current observations when interpreted via a Friedmann-Lema\^{i}tre-Robertson-Walker (FLRW) universe governed by general relativity point to a universe that is flat or very close to it, modified theories of gravity or inhomogeneous cosmological models may in fact require curvature to fit observations \cite{r42a,r43a,r44a}. In order to perform the analysis, we expand the framework of modified growth (MG) equations and parameters. We also expand the corresponding numerical framework described in \cite{r40} to include curved models and apply these changes to the publicly available code, {\it {I}ntegrated {S}oftware {i}n {T}esting {G}eneral {R}elativity} (\texttt{ISiTGR}) \cite{ISITGRweb}. To constrain the MG parameters while allowing for spatial curvature, we will then use current observations including: the WMAP7 temperature and polarization spectra (CMB) \cite{r42}; the matter power spectrum (MPK) from the Sloan Digital Sky Survey (SDSS) DR7 \cite{r43}; the Integrated Sachs-Wolfe (ISW)-galaxy cross-correlations \cite{r44,r45}; the refined HST-COSMOS weak-lensing tomography \cite{r46}; WiggleZ BAO measurements \cite{r47}; the supernovae compilation of the Supernovae Cosmology Project (SCP)  \cite{r48} and references of other compiled supernovae (SN) therein; the prior on $H_0=74.2\pm 3.6$ km/s/Mpc given by Ref. \cite{r49}; and finally, a prior on the age of the Universe (AGE) $10\,$Gyrs$<$AGE$<20\,$Gyrs.  We will explore the effect this has on the constraints compared to the flat case.  Next, we will analyze the correlations between the curvature parameter $\Omega_k$ and the MG parameters.  Finally, we will explore what effect assuming a flat model when curvature is present will have on the best-fit MG parameters, to see if a bias is introduced by such an assumption and discuss how this bias is related to the correlation coefficients. We conclude in the last section. 
%
%

%
\section{Growth Equations Including Spatial Curvature in General Relativity}
The perturbed FLRW metric written in the general conformal Newtonian gauge is given by
\be
ds^2=a(\tau)^2[-(1+2\psi)d\tau^2+(1-2\phi)\gamma_{ij}dx^idx^j],
\label{eq:FLRWNewt}
\ee
where $\phi$ and $\psi$ are scalar potentials describing the scalar mode of the metric perturbations, $\tau$ is conformal time, $a(\tau)$ is the scale factor normalized to one today, and the $x_i$'s are the comoving coordinates. $\gamma_{ij}$ is the 3-metric, which can be written in coordinates $(x,\,y,\,z)$ as
\be
\gamma_{ij} = \delta_{ij}\left[1+\frac{K}{4}\left(x^2+y^2+z^2\right)\right]^{-2},
\label{eq:3met}
\ee
where $K=-\Omega_k\mathcal{H}_0^2$ is the spatial curvature, and $\mathcal{H}_0$ is the Hubble parameter today.

As discussed in \cite{r50,r51}, when working in a non-flat universe the Fourier modes are generalized as eigenfunctions, $G$, of the Laplacian operator such that
\be 
\nabla^2G(\vec{k},\vec{x})= -k^2G(\vec{k},\vec{x}).
\ee
In our analysis we expand perturbations in terms of $G$ and its spatial covariant derivatives (denoted by ${_|}$) as seen in \cite{r51}.

Two very useful equations can be obtained using the first-order perturbed Einstein equations.  These equations relate the scalar potentials to one another, as well as both the gauge-invariant, rest-frame overdensity, $\Delta_i$, and the shear stress, $\sigma_i$ (where $_i$ denotes a particular matter species).  First combining the time-space and time-time equations gives the Poisson equation, while the traceless, space-space component of the equations gives a relation between the two metric potentials.  Explicitly, these equations are
\bea
\left(k^2-3K\right)\phi  &=&-4\pi G a^2\sum_i \rho_i \Delta_i
\label{eq:Poisson}\\
k^2(\psi-\phi) &=& -12 \pi G a^2\sum_i \rho_i(1+w_i)\sigma_i,
\label{eq:2ndEin}
\eea
where $\rho_i$ is the density for matter species $i$.  

The quantity $\Delta_i$ is important because its evolution can be used to describe the growth of inhomogeneities.  It is defined in terms of the fractional overdensity, $\delta_i=\delta \rho_i/\bar{\rho}$, and the heat flux, $q_i$ ($q_i$ is related to the divergence of the peculiar velocity, $\theta_i$, by $\theta_i=\frac{k\ q_i}{1+w_i}$), as
\be
\Delta_i = \delta_i +3\mathcal{H}\frac{q_i}{k},
\label{eq:Delta}
\ee 
where $\mathcal{H} =\dot{a}/a$ is the Hubble factor in conformal time. By considering mass-energy conservation, the evolution of $\Delta$ for an uncoupled fluid species or its mass average for all the fluids is given by \cite{r52,r40}
\be
\dot{\Delta} = 3(1+w)\left(\dot{\phi}+\mathcal{H}\psi\right)+3\mathcal{H}w\Delta -\left[k^2+3\left(\mathcal{H}^2-\dot{\mathcal{H}}\right)\right]\frac{q}{k}-3\mathcal{H}(1+w)\sigma,
\label{eq:Deltadot}
\ee
where $w=p/\rho$ is the equation of state of the fluid.  
Combining Eqs. (\ref{eq:Poisson}), (\ref{eq:2ndEin}), and (\ref{eq:Deltadot}) along with the evolution equations for $a(\tau)$, the growth history of structures in the universe can be fully described.
%
\section{Modifications to the Growth Equations For Spatially Curved Models}

\subsection{Modified growth equations in the conformal Newtonian gauge}

Recently, one of the major routes to testing general relativity has been parametrizing both the relation between the two metric potentials $\phi$ and $\psi$ in the perturbed FLRW metric (an inequality in this relation has been called \emph{gravitational slip} by \cite{r17}) as well as modifications to Poisson's equation, Eq. (\ref{eq:Poisson}). Examples of this approach can be seen in \cite{r17,r53,r35, r33, r54, r34, r36, r37, r38, r55, r39, r40, r56, r57, r41} to name a few.  To date, though, explorations of these modifications have only been performed in flat spacetimes.    We now focus on extending the modified growth equations seen in, for example, \cite{r33,r40} to non-flat cases.

Extending the parametrized modifications of the growth equations, (\ref{eq:Poisson}) and (\ref{eq:2ndEin}), proposed by \cite{r33} to non-flat models gives
\bea
\left(k^2-3K\right)\phi  &=& -4\pi G a^2\sum_i \rho_i \Delta_i \,  Q
\label{eq:PoissonMod}\\
k^2(\psi-R\,\phi) &=& -12 \pi G  a^2\sum_i \rho_i(1+w_i)\sigma_i \, Q,
\label{eq:Mod2ndEin}
\eea
where $Q$ and $R$ are the modified growth parameters (MG parameters).  A modification to the Poisson equation is quantified by the parameter $Q$, while the gravitational slip is quantified by the parameter $R$ (at late times, when anisotropic stress is negligible, $\psi= R\phi$).  As discussed in our earlier paper \cite{r40} we use the parameter $\mathcal{D} = Q(1+R)/2$ instead of $R$ not only to avoid a strong degeneracy between $Q$ and $R$, but also to have a parameter which can be directly probed by observations.  To obtain a modified growth equation written in terms of only $Q$ and $\mathcal{D}$, we combine Eqs. (\ref{eq:PoissonMod}) and (\ref{eq:Mod2ndEin}), giving
\be
k^2(\psi+\phi) = \frac{-8\pi G a^2}{1-3K/k^2}\sum_i \rho_i \Delta_i \,\mathcal{D} \, -12 \pi G  a^2\sum_i \rho_i (1+w_i)\sigma_i \, Q.
\label{eq:PoissonModSum}
\ee

\subsection{Modified equations in the synchronous gauge}
In order to perform the tests, we must implement the MG framework into numerical codes that allow comparisons to the data and calculations of parameter constraints.
This is done by an extended version, with the inclusion of spatial curvature, of the publicly available package \texttt{ISiTGR} which is an integrated set of modified modules for the publicly available codes \texttt{CosmoMC} \cite{r58} and \texttt{CAMB} \cite{r59}. \texttt{CAMB} is used to calculate the various CMB  anisotropy spectra ($C_\ell^{TT}$, $C_\ell^{TE}$, $C_\ell^{EE}$, $C_\ell^{BB}$) as well as the three-dimensional matter power spectrum $P_\delta(k,z)$ all of which are very powerful in constraining both the growth history of structure in the universe as well as the expansion history of the universe, and thus MG parameters.

The package \texttt{CAMB} is written in the synchronous gauge, where the perturbed FLRW metric is written as:
\be
ds^2=a(\tau)^2[-d\tau^2+(\gamma_{ij}+h_{ij})dx^idx^j].
\label{eq:FLRWSync}
\ee
Thus, instead of using the metric potentials $\phi$ and $\psi$ of the conformal Newtonian gauge, it uses the metric potentials $h$ and $\eta$ consistent with the notation of \cite{r51}. These new perturbation quantities are related to the trace ($h$) and traceless ($h+6\eta$) part of the total metric perturbation $h_{ij}$.  Explicitly, by writing $h_{ij}$ expanded in terms of $G$, described above, for a single mode we have \cite{r51}
\be
h_{ij} = \frac{h}{3}\gamma_{ij}G+(h+6\eta)(k^{-2}G_{|ij}+\frac{1}{3}\gamma_{ij}G).
\ee
Now combining the perturbed Einstein's equations in the same way as discussed to get Eqs. (\ref{eq:Poisson}) and (\ref{eq:2ndEin}), we have for the synchronous gauge \cite{r60}
\bea
\left(k^2-3K\right)(\eta-\mathcal{H}\alpha)  &=&-4\pi G a^2\sum_i \rho_i \Delta_i,
\label{eq:PoissonSync}\\
k^2(\dot{\alpha}+2\mathcal{H}\alpha-\eta) &=& -12 \pi G a^2\sum_i \rho_i(1+w_i)\sigma_i,
\label{eq:2ndEinSync}
\eea
where $\alpha = (h+6\eta)/2k^2$. 

Armed with the knowledge that both $\Delta_i$ and $\sigma_i$ are invariant between these two gauges and quickly comparing Eqs. (\ref{eq:Poisson}) and (\ref{eq:PoissonSync}) as well as (\ref{eq:2ndEin}) and (\ref{eq:2ndEinSync}), we can see the metric potentials in the two gauges are related to one another by:
\bea
\phi & = & \eta -\mathcal{H}\alpha,
\label{eq:Phisync} \\
\psi &= & \dot{\alpha}+\mathcal{H} \alpha,
\label{eq:Psisync}
\eea 

\texttt{CAMB} evolves the metric potential $\eta$ (this is done by evolving the quantity $k\eta$).  For each matter species, the code also evolves: the matter perturbations, $\delta_i$; the heat flux, $q_i$; and the shear stress $\sigma_i$ according to the synchronous gauge evolution equations given in \cite{r52}. The code has been written in such a way that the evolution of all other variables is changed simply by adjusting the evolution of the metric potential $\eta$.  So the first step to modifying \texttt{CAMB} to be consistent with the modified growth Eqs. (\ref{eq:PoissonMod}) and (\ref{eq:PoissonModSum}) is to derive an equation for the evolution of $\eta$ consistent with those equations.  We begin by subbing (\ref{eq:Phisync}) into (\ref{eq:PoissonMod}) and taking the time derivative.  This gives
\be
k^2\dot{\eta} = -\frac{1}{2 K_{f1} } \sum_i\left[\tilde{\rho_i}(a)(\dot{Q}\Delta_i + Q\dot{\Delta}_i)+Q\Delta_i\frac{d}{d\tau}\tilde{\rho_i}(a)\right] +\dot{\mathcal{H}}k^2\alpha +\mathcal{H}k^2\dot{\alpha},
\label{eq:ekd1}
\ee
where we have used the condensed notation, $\tilde{\rho_i}(a) = 8\pi G a^2\rho_i$ and $K_{f1}  = 1-3K/k^2$. Next an expression for $\dot{\alpha}$ is obtained by substituting Eqs. (\ref{eq:PoissonMod}) and (\ref{eq:Psisync}) into Eq. (\ref{eq:PoissonModSum}), yielding
\be 
\dot{\alpha} = -\mathcal{H}\alpha - \frac{1}{2k^2}\sum_i\tilde{\rho_i}(a)\left[\frac{2\mathcal{D}-Q}{K_{f1} }\Delta_i + 3Q(1+w_i)\sigma_i \right].
\label{eq:dotalpha}
\ee
Now subbing the time derivative of $\rho_i$ (obtained from matter conservation) as well as Eqs. (\ref{eq:Deltadot}) and (\ref{eq:dotalpha}) into (\ref{eq:ekd1}) gives:
\bea
k^2\dot{\eta}=\frac{-1}{2\, K_{f1}}\sum_i\tilde{\rho_i}(a)\Bigg\{ \dot{Q}\Delta_i - \mathcal{H}Q\Delta_i + 3Q(1+w_i)\left(\dot{\phi} + \mathcal{H}\psi\right)  -3\mathcal{H}(1+w_i)\sigma_i \nonumber \\
- Q f_1\frac{q_i}{k} +2\mathcal{H}\mathcal{D}\Delta_i -\mathcal{H}Q\Delta_i +3\mathcal{H}Q\, K_{f1}(1+w_i)\sigma_i \Bigg\} -(\mathcal{H}^2-\dot{\mathcal{H}})k^2\alpha,
\label{eq:ekd2}
\eea
where
\be
f_1  =  k^2 + 3 (\mathcal{H}^2-\dot{\mathcal{H}}).
\ee
Next we substitute Eq. (\ref{eq:Psisync}) and the time derivative of  Eq. (\ref{eq:Phisync}) into Eq. (\ref{eq:ekd2}), giving
\bea
k^2\dot{\eta}=\frac{-1}{2\, K_{f1}}\sum_i\tilde{\rho_i}(a)\Bigg\{ \dot{Q}\Delta_i - \mathcal{H}Q\Delta_i + 3Q(1+w_i)\left(\dot{\eta} + (\mathcal{H}^2-\dot{\mathcal{H}})\alpha\right) -3\mathcal{H}(1+w_i)\sigma_i \nonumber \\
- Q f_1\frac{q_i}{k} +2\mathcal{H}\mathcal{D}\Delta_i -\mathcal{H}Q\Delta_i +3\mathcal{H}Q\, K_{f1}(1+w_i)\sigma_i \Bigg\} -(\mathcal{H}^2-\dot{\mathcal{H}})k^2\alpha.
\label{eq:ekd3}
\eea

Since the variables in \texttt{CAMB} are evolved in the synchronous gauge, we must make sure that all the quantities in the $\dot{\eta}$ are the synchronous gauge quantities.  We know that $\Delta$ and $\sigma$ are gauge invariant, so we need not be concerned about them.  However, $q_i$ in the above equations is still a conformal Newtonian gauge-quantity.  Using Eq. 27b of \cite{r52} (first converting from $\theta$ to $q$ as described above) the transformation of $q_i$ between these two gauges is written as
\be
q_i^{(c)}=q_i^{(s)}+(1+w_i)k\alpha.
\label{eq:transq}
\ee 
Upon subbing Eq. (\ref{eq:transq}) into Eq. (\ref{eq:ekd3}) and collecting the terms we finally arrive at an equation for $\dot{\eta}$:
\bea
\dot{\eta}=\frac{-1}{2\,K_{f1} \,f_Q}\sum_i\tilde{\rho_i}(a)\Bigg\{  \left(2\mathcal{H}\left[\mathcal{D}-Q\right]+\dot{Q}\right)\Delta_i -Q(1+w_i)k^2\alpha  \nonumber \\
 -  Q f_1\frac{q_i}{k}+3\mathcal{H}Q (K_{f1} -1) (1+w_i)\sigma_i \Bigg\}-\frac{\mathcal{H}^2-\dot{\mathcal{H}}}{f_Q}k^2\alpha ,
\label{eq:ekdfin}
\eea
with:
\be
f_Q  = k^2 + \frac{3Q}{2K_{f1}}\sum_i\tilde{\rho_i}(1+w_i)
\ee

Now as discussed in \cite{r40} the time derivative of the sum of the Newtonian metric potentials, $\dot{\phi}+\dot{\psi}$, goes into evaluating the contribution of the ISW effect in the CMB temperature anisotropy. This quantity needs to be consistent with the modified growth equation (\ref{eq:PoissonModSum}).  Thus, simply taking the time derivative of (\ref{eq:PoissonModSum}) and subbing in for $\dot{\Delta}$ and $\dot{\tilde{\rho}}_i$ gives:
\bea
\dot{\phi}+\dot{\psi} = \frac{1}{k^2}\sum_i\tilde{\rho_i}(a)\Bigg\{\left[\left((1+3w_i)Q+\frac{2\mathcal{D}}{K_{f1}}\right)\mathcal{H}-\dot{Q}\right]\frac{3(1+w_i)\sigma_i}{2} -\frac{3Q(1+w_i)\dot{\sigma}_i}{2} \\ \nonumber
+\frac{1}{K_{f1}}\left[(\mathcal{D}\mathcal{H}-\dot{\mathcal{D}})\Delta_i+\mathcal{D}(1+w_i)\left(k^2\alpha-3\dot{\eta}\right)+ \mathcal{D} f_1\frac{q_i}{k}\right] \Bigg\}.
\eea

\subsection{Modified power spectra}
In the initial release of \texttt{ISiTGR} for both weak lensing tomography and ISW-galaxy cross-correlation measurements, the output 3D power spectrum $P_{\delta,\delta}(k,z)$ was calculated by \texttt{CAMB} and then used to calculate the power spectra needed for lensing, $P_{\Phi,\Phi}(k,z)$, and the power spectrum need for ISW-galaxy cross-correlations, $P_{\delta,2\dot{\Phi}}(k,z)$, where $\Phi \equiv \frac{\phi+\psi}{2}$ and $A$ and $B$ in $P_{A,B}$ refer to the transfer functions used in the calculation of a given power spectrum.  In this updated version of \texttt{ISiTGR} these power spectra are calculated directly in \texttt{CAMB}.  A detailed description of the weak lensing and ISW-galaxy cross-correlation likelihoods can be found in \cite{r40}, but we will review the most relevant equations here.   

The lensing cross power spectrum between redshift bins $k$ and $l$ is given by \cite{r61}
\be
P_\kappa^{kl}(\ell) = \int^{\chi_h}_{0}  d\chi \frac{g_k(\chi)g_l(\chi)}{a^2(\chi)}\, P_{\Phi,\Phi} \Big(\frac{\ell}{f_K(\chi)},\chi \Big),
\label{eq:LCPS}
\ee
with comoving radial distance $\chi$, comoving distance to the horizon $\chi_h$,comoving angular diameter distance $f_K(\chi)$, and where we have absorbed the usual extra terms into the $P_{\Phi,\Phi}$. The weighted geometric lens-efficiency factor for the $k^{th}$ bin $g_k(\chi)$ given by 
\be 
g_k(\chi)  \equiv \int^{\chi_h}_{\chi} d\chi' p_k(\chi') \frac{f_K(\chi'-\chi)}{f_K(\chi')},
\label{eq:LensEff}
\ee
corresponding to the galaxy redshift distributions $p_k$.

The ISW-galaxy cross power spectrum is written as \cite{r44}:
\be
C_\ell^{gT} = \frac{T_{CMB}}{(\ell +1/2)^2}\int dz b(z)\Pi(z) P_{\delta,2\dot{\Phi}}\left(\frac{\ell+1/2}{f_K(\chi(z))},z\right),
\label{eq:ISWgalSpec}
\ee
where $b(z)$ is the galaxy bias, $\Pi(z)$ is the normalized selection function, and again we have absorbed the usual extra terms into the $P_{\delta,2\dot{\Phi}}$.

\begin{center}
\begin{table}[b]
\begin{tabular}{|c|c|c|c|c|c|c|c|}\hline
\multicolumn{8}{|c|}{{ \bfseries Correlation coefficients between $\Omega_k$ and the MG parameters}}\\ \hline
\multicolumn{8}{|c|}{{MG parameters evolved using traditional binning}}\\ \hline
$Q_1$&$Q_2$&$Q_3$&$Q_4$&$\mathcal{D}_1$&$\mathcal{D}_2$& $\mathcal{D}_3$&$\mathcal{D}_4$\\ \hline
\,-0.1658 \,& \, 0.0108 \,  & \,  -0.1114  \, &  \, -0.1872 \,  &  \, -0.0752 \,  & \,  -0.3606 \,  & \,  -0.0327  \, & \,  -0.0988 \,  \\ \hline
\multicolumn{8}{|c|}{}\\ \hline
\multicolumn{8}{|c|}{{ MG parameters evolved using hybrid binning}}\\ \hline
$Q_1$&$Q_2$&$Q_3$&$Q_4$&$\mathcal{D}_1$&$\mathcal{D}_2$& $\mathcal{D}_3$&$\mathcal{D}_4$\\ \hline
\,-0.0662\, &\,-0.0840\, &\,-0.1258\, &\,0.1365\, &\,0.0121\, &\,-0.2480\, &\,0.1025\, &\,-0.2316\, \\ \hline
\end{tabular}
\\
\begin{tabular}{|c|c|c|c|}\hline 
\multicolumn{3}{|c|}{}\\ \hline
\multicolumn{3}{|c|}{{ MG parameters evolved using the functional form}}\\ \hline
$Q_0$&$\mathcal{D}_0$&$R_0$\\ \hline
\,\,\,\,\,\, \,\,\,-0.0166 \,\,\,\,\,\, \,\,\, & \,\,\,\,\,\, \,\,\,0.1428 \,\,\,\,\,\, \,\,\,& \,\,\, \,\,\,0.0400 \,\,\, \,\,\, \\ \hline
\end{tabular} 
\caption{\label{table:CurCorr}
We list the correlation coefficients between $\Omega_k$ and the various MG parameters for the current observed data.  We see that the MG parameters for all evolution methods are somewhat correlated with $\Omega_k$.  The functional form method shows overall the least amount of correlation between the MG parameters and the two cosmological parameters.  The hybrid binning is the next least correlated of the methods, while parameters from the traditional binning method of evolution show the most amount of correlation with $\Omega_k$.
}
\end{table}
\end{center}
\section{Results}

\subsection{Modified growth and cosmological parameters used}
For all results, in addition to the curvature parameter $\Omega_k$ and the MG parameters we vary the six core cosmological parameters: $\Omega_bh^2$ and $\Omega_c h^2$, the baryon and cold dark matter physical density parameters, respectively; $\theta$, the ratio of the sound horizon to the angular diameter distance of the surface of last scattering; $\tau_{rei}$, the reionization optical depth; $n_s$, the spectral index; and $\ln10^{10} A_s$, the amplitude of the primordial power spectrum.

We use three different parametrizations of the MG parameters:  two scale-dependent methods including a traditional binning method and a hybrid evolution method, both of which were discussed in our previous work \cite{r40}, and a scale independent method using a functional first introduced by \cite{r33}:  
\begin{enumerate}  
\item{For traditional binning both redshift, $z$, and scale, $k$, are binned in two bins creating a total of four bins.  The $z$-bins are $0<z\le1$ and $1<z\le 2$ while for $z>2$ GR is assumed valid.  The $k$-bins are simply $k\le0.01$ and $k>0.01$.  This binning can be described functionally as 
\bea
Q(k,a) &=&\frac{1}{2}\big(1 + Q_{z_1}(k)\big)+\frac{1}{2}\big(Q_{z_2}(k) - Q_{z_1}(k)\big)\tanh{\frac{z-1}{0.05}}+\frac{1}{2}\big(1 - Q_{z_2}(k)\big)\tanh{\frac{z-2}{0.05}},\label{eq:ZBinEvo}\\ \nonumber
\mathcal{D}(k,a) &=&\frac{1}{2}\big(1 + \mathcal{D}_{z_1}(k)\big)+\frac{1}{2}\big(\mathcal{D}_{z_2}(k) - \mathcal{D}_{z_1}(k)\big)\tanh{\frac{z-1}{0.05}}+\frac{1}{2}\big(1 - \mathcal{D}_{z_2}(k)\big)\tanh{\frac{z-2}{0.05}},
\eea
with
\bea
Q_{z_1}(k) &=& \frac{1}{2}\big(Q_2+Q_1\big)+\frac{1}{2}\big(Q_2-Q_1\big)\tanh{\frac{k-0.01}{0.001}},
\label{eq:kBinQ} \\ \nonumber
Q_{z_2}(k) &=& \frac{1}{2}\big(Q_4+Q_3\big)+\frac{1}{2}\big(Q_4-Q_3\big)\tanh{\frac{k-0.01}{0.001}},\\
\mathcal{D}_{z_1}(k) &=& \frac{1}{2}\big(\mathcal{D}_2+\mathcal{D}_1\big)+\frac{1}{2}\big(\mathcal{D}_2-\mathcal{D}_1\big)\tanh{\frac{k-0.01}{0.001}},
\label{eq:kBinD} \\ \nonumber
\mathcal{D}_{z_2}(k) &=& \frac{1}{2}\big(\mathcal{D}_4+\mathcal{D}_3\big)+\frac{1}{2}\big(\mathcal{D}_4-\mathcal{D}_3\big)\tanh{\frac{k-0.01}{0.001}},\\
\eea
This gives 8 MG parameters, $\mathcal{D}_i$ and $Q_i$, $i=1,2,3,4$ which are varied.}

\item{The redshift evolution hybrid evolution method is identical to that of traditionally binning ie Eq.(\ref{eq:ZBinEvo}).  The evolution in scale, however, is described by a monotonic function written as \cite{r40},
\bea
Q_{z_1}(k) &=& Q_1 e^{-\frac{k}{0.01}}+Q_2(1-e^{-\frac{k}{0.01}}), 
\label{eq:kHybridQ} \\ \nonumber
Q_{z_2}(k) &=& Q_3 e^{-\frac{k}{0.01}}+Q_4(1-e^{-\frac{k}{0.01}}),\\
\mathcal{D}_{z_1}(k) &=& \mathcal{D}_1 e^{-\frac{k}{0.01}}+\mathcal{D}_2(1-e^{-\frac{k}{0.01}}),
\label{eq:kHybridD} \\ \nonumber
\mathcal{D}_{z_2}(k) &=& \mathcal{D}_3 e^{-\frac{k}{0.01}}+\mathcal{D}_4(1-e^{-\frac{k}{0.01}}).
\eea
This provides an exponential transition in scale between two parameter values in each redshift bin.  Again a total of 8 MG parameters are varied, $\mathcal{D}_i$ and $Q_i$, $i=1,2,3,4$}

\item{For the functional form method we vary the MG parameters $Q$ and $R$ according to \cite{r33}
\bea
Q(a) &=& \left(Q_0-1\right)a^s +1, 
\label{eq:QFunc}\\
R(a) &=& \left(R_0 -1\right)a^s +1.
\label{eq:RFunc}
\eea
The parameters $Q_0$ and $R_0$ are the present day values of the MG parameters which go to 1 at early times.  The parameter $s$ describes the time dependence of the MG parameters and is marginalized over.  To alleviate a known degeneracy in the parameter space between $Q_0$ and $R_0$, we vary $Q_0$ and $\mathcal{D}_0=Q_0(1+R_0)/2$ inferring $R_0$ from the other two parameters.}
\end{enumerate}  

\begin{samepage}
\begin{figure}[t]
\begin{center}
\begin{tabular}{|c|}
\hline 
{\includegraphics[width=4.33in,height=2.5in,angle=0]{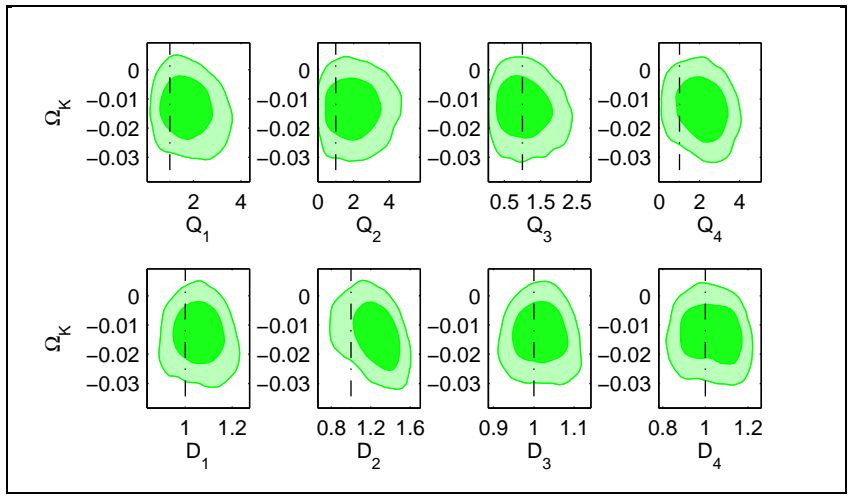}} \\ \hline
\end{tabular}
\caption{\label{figure:CorrBin}We plot the 2D confidence contours for $\Omega_k$ and the MG parameters $Q_i$ and $\mathcal{D}_i$, $i=1,2,3,4$ from using traditional bins for $k$ and $z$.  As seen in Table \ref{table:CurCorr} this evolution method has the most correlation between the MG parameters and $\Omega_k$ of the three evolution methods.} 
\end{center}
\end{figure}   
\begin{figure}[t]
\begin{center}
\begin{tabular}{|c|}
\hline 
{\includegraphics[width=4.33in,height=2.5in,angle=0]{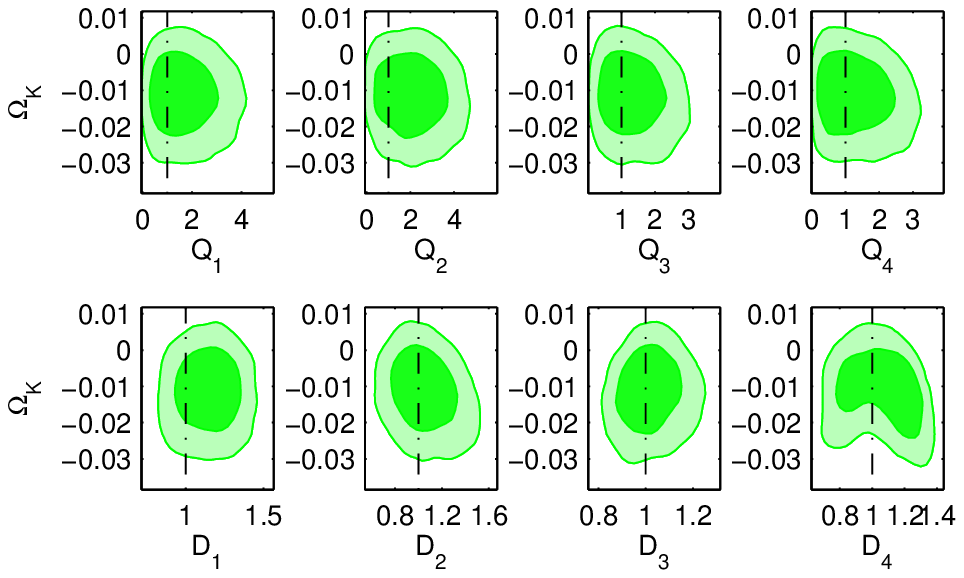}}\\ \hline
\end{tabular}
\caption{\label{figure:CorrHybrid}
We plot the 2D confidence contours for $\Omega_k$ and the MG parameters $Q_i$ and $\mathcal{D}_i$, $i=1,2,3,4$ from the hybrid method to evolve the MG parameters.  As seen in Table \ref{table:CurCorr} this evolution method has a moderate amount of correlation between the MG parameters and $\Omega_k$.} 
\end{center}
\end{figure}

\begin{figure}[t]
\begin{center}
\begin{tabular}{|c|}
\hline 
{\includegraphics[width=3.25in,height=1.25in,angle=0]{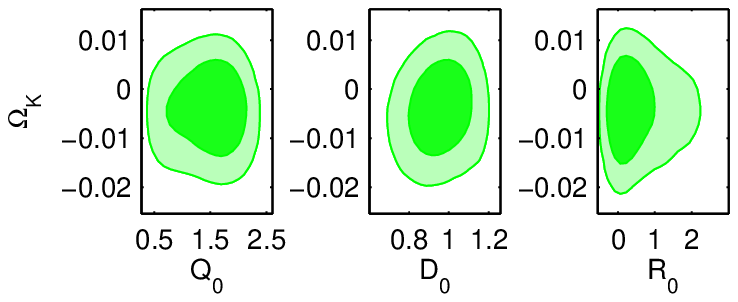}} \\
\hline   
\end{tabular}
\caption{\label{figure:CorrFunc}
We plot the 2D confidence contours for $\Omega_k$ and the MG parameters $Q_0$ and $\mathcal{D}_0$, from using the functional form to evolve the MG parameters.  As seen in Table \ref{table:CurCorr} this evolution method overall has the least amount of correlations between the MG parameters and the $\Omega_k$ of the three evolution methods.} 
\end{center}
\end{figure}
\end{samepage}

\subsection{Results from current data and correlations between $\Omega_k$ and the MG parameters}
Here we study the correlation coefficients between the MG parameters and $\Omega_k$.   We choose to focus on this parameter because a quick examination of Eqs. (\ref{eq:PoissonMod}) and (\ref{eq:PoissonModSum}) shows that the parameters should be correlated. Indeed, using Eq. (\ref{eq:PoissonModSum}) in particular shows how this correlation is expected.  The first term on the right-hand side of this equation has $K$, the spatial curvature, in the denominator while the MG parameter $\mathcal{D}$ is in the numerator.  An increased $K$ and thus decreased $\Omega_k$ will (for a given range of wave number $k$) be balanced by a decreased $\mathcal{D}$.  This is expected since an increased $K$ (decreased $\Omega_k$) makes the denominator smaller and thus the whole first term larger.  A decreased $\mathcal{D}$ will balance this effect.  The same effect can be seen for the MG parameter $Q$ if we rearrange Eq. (\ref{eq:PoissonMod}) in a form similar to Eq. (\ref{eq:PoissonModSum}).  The presence of this degenerate effect between the MG parameters and $\Omega_k$ is the reason we choose to focus on the said correlation in this paper.  No other correlations are as obvious through an analytical examination.

We first discuss correlation coefficients that were obtained from constraints using the latest cosmological data including; the WMAP7 temperature and polarization spectra (CMB) \cite{r42}, the matter power spectrum (MPK) from the Sloan Digital Sky Survey (SDSS) DR7 \cite{r43}, the ISW-galaxy cross-correlations \cite{r44,r45}, the refined HST-COSMOS weak-lensing tomography \cite{r46}, WiggleZ BAO measurements \cite{r47}, and the supernovae compilation of the Supernovae Cosmology Project (SCP) \cite{r48} and references of other compiled supernovae (SN) therein.  We also use the prior on $H_0=74.2\pm 3.6$ km/s/Mpc given by \cite{r49}, and a prior on the age of the Universe (AGE) $10\,$Gyrs$<$AGE$<20\,$Gyrs. The constraints on the MG parameters we obtain are all consistent with general relativity at the 95\% confidence level.  

In Table \ref{table:CurCorr} we list the correlations between $\Omega_k$ and the MG parameters.  We have used the standard definition for the correlation coefficient of two parameters, $p_x$ and $p_y$:
\be
Corr(p_x,p_y)=\frac{Cov(p_x,p_y)}{\sigma({p_x})\sigma({p_y})},
\ee
where the covariance of the two parameters is $Cov(p_x,p_y)$ and their respective standard deviations are $\sigma({p_x})$ and $\sigma({p_y})$.  
We further illustrate these correlations by plotting the 2D confidence contours for the functional form, traditional binning, and hybrid evolution methods in Figs. \ref{figure:CorrBin}, \ref{figure:CorrHybrid}, and \ref{figure:CorrFunc}, respectively.

The existence of non-negligible correlations between $\Omega_k$ and the MG parameters indicates that curvature may possibly produce an apparent deviation from GR in the MG parameter space if the universe is assumed to be flat.  To explore this possibility further, in the next section we generate future cosmological data with different curvature values and see how much the MG parameter constraints are affected by assuming a spatially flat background when, in fact, the background is curved.

\begin{figure}[t]
\begin{center}
\begin{tabular}{|c|}
\hline 
{\includegraphics[width=4.33in,height=1.25in,angle=0]{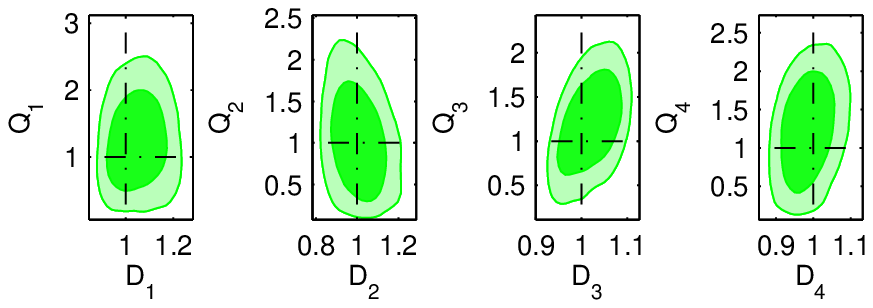}}\\ \hline
{\includegraphics[width=4.33in,height=1.25in,angle=0]{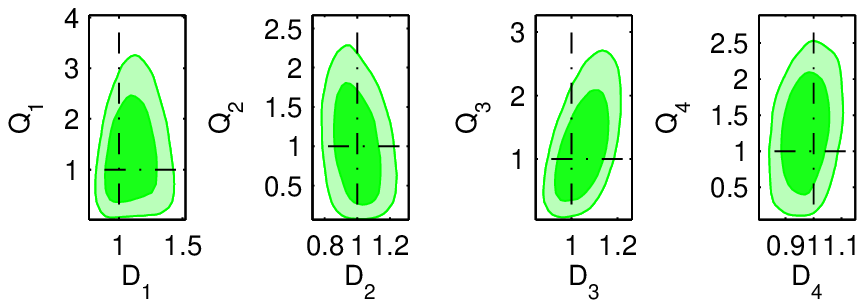}}\\ \hline
\end{tabular}
\\
\begin{tabular}{|c|}
\hline
{\includegraphics[width=2.16in,height=1.25in,angle=0]{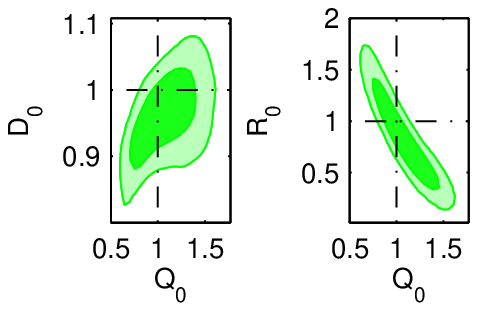}} \\
\hline   
\end{tabular}
\caption{\label{figure:F_kp01} 
Plotted are the 68\% and 95\% 2D confidence contours for the MG parameters when a spatially flat model is used but the actual underlying universe has $\Omega_k=0.01$.
TOP: Confidence contours for the MG parameters $Q_i$ and $\mathcal{D}_i$, $i=1,..4$ when the traditional binning method is used.  All parameter values are pulled to smaller values.  MIDDLE: Confidence contours for the MG parameters $Q_i$ and $\mathcal{D}_i$, $i=1,..4$ when the hybrid evolution method is used.  Most of the parameter contours are pulled to smaller values. BOTTOM:  Confidence contours for the MG parameters $Q_0$ and $\mathcal{D}_0$ and $R_0$ when the functional form binning method is used.  The $Q_0-\mathcal{D}_0$ contour is pulled noticeably toward smaller parameter values.} 
\end{center}
\end{figure}
\subsection{Constraints from simulated future data and the effect of assuming a spatially flat model on curved ones}
To explore the effect of assuming a flat background when the actual cosmology is curved (even slightly) we generate fiducial data with uncertainties achievable by future higher precision observations. It is worth noting here that the constraints obtained in this section are not meant to represent the constraining power of  future observations, but rather to illustrate how the assumption of spatial flatness may affect constraints on MG parameters.  The fiducial data sets are produced using a curved $\Lambda$CDM model so all of the MG parameters are set to their GR values during data generation.  For each fiducial data set, we generated data for the following observations: CMB (Plank-like), matter power spectrum, weak-lensing, ISW-galaxy cross-correlations, Type Ia supernovae, and baryon acoustic oscillations.  The CMB data was produced using \texttt{CosmoMC} and we used the included \texttt{PYTHON} script to add noise and make it compatible with the included Plank likelihood which was used during the fits.  For ISW-galaxy cross-correlations, weak lensing and matter power spectrum data, the current galaxy distribution and window files were used.  Random noise up to 5\%  of the true data value was then added to the data. For likelihood calculations uncertainties were reduced by a factor of 3 compared to current observations.  BAO observations were given random noise of up to 5\% while supernova data points were allowed random noise up to 2.5 times current uncertainties.  Uncertainties in these two data sets were not improved compared to currently available data, primarily because they probe only the expansion history of the universe and so are not important in constraining the MG parameters.  For all fits, in addition to varying the MG parameters, the six core cosmological parameters are varied as well.  A null test, where $\Omega_k$ is also varied, was performed on all fiducial data sets generated and the results of these fits in all cases recover the parameter values near those used to generate the fiducial data.  This ensures that if indeed non-GR constraints are obtained for the MG parameters when we assume a flat universe, curvature is the cause and not a flaw in the production of the fiducial data.

\begin{figure}[t]
\begin{center}
\begin{tabular}{|c|}
\hline 
{\includegraphics[width=4.33in,height=1.25in,angle=0]{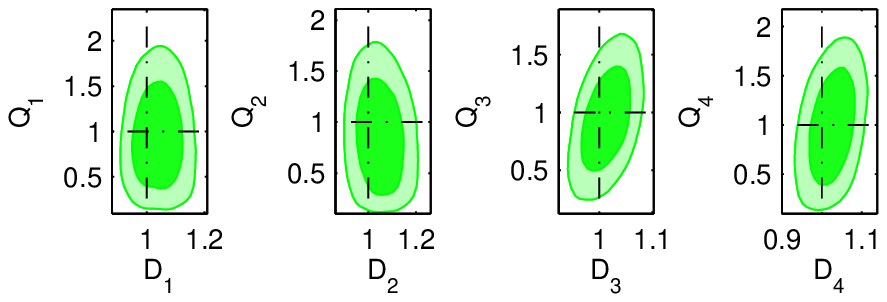}}\\ \hline
{\includegraphics[width=4.33in,height=1.25in,angle=0]{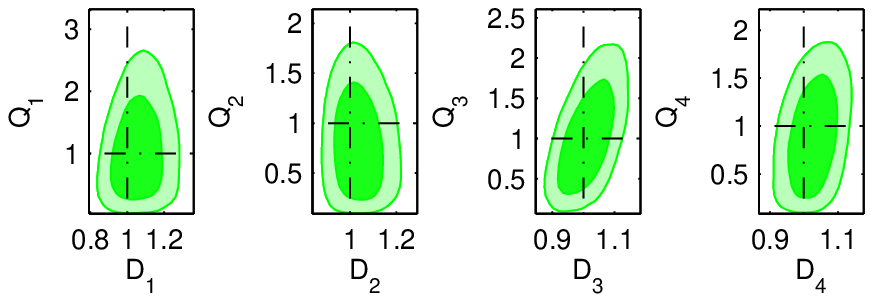}}\\ \hline
\end{tabular}
\\
\begin{tabular}{|c|}
\hline
{\includegraphics[width=2.16in,height=1.25in,angle=0]{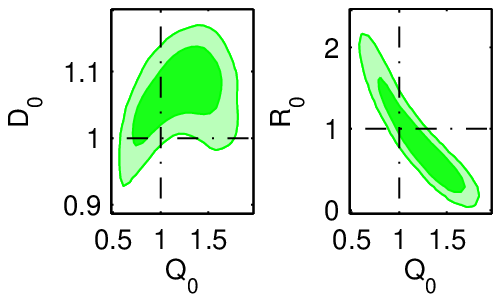}} \\
\hline   
\end{tabular}
\caption{\label{figure:F_knp02}
 Plotted are the 68\% and 95\% 2D confidence contours for the MG parameters when a spatially flat model is used but the actual underlying universe has $\Omega_k=-0.02$
TOP: Confidence contours for the MG parameters $Q_i$ and $\mathcal{D}_i$, $i=1,..4$ when the traditional binning method is used.  All parameter values are pulled to larger values.   MIDDLE: Confidence contours for the MG parameters $Q_i$ and $\mathcal{D}_i$, $i=1,..4$ when the hybrid evolution method is used.  Most of the parameter contours are pulled to larger values. BOTTOM:  Confidence contours for the MG parameters $Q_0$ and $\mathcal{D}_0$ and $R_0$ when the functional form binning method is used.  The $Q_0-\mathcal{D}_0$ contour is pulled noticeably toward larger parameter values.}  
\end{center}
\end{figure}
Two curved fiducial models were initially tested: one negatively curved model with $\Omega_k=0.01$ and one positively curved model with $\Omega_k=-0.02$. These values were picked because they are the rounded WMAP7 95\% confidence limits on $\Omega_k$.  The constraints on the MG parameters for these data sets is plotted in Figs. \ref{figure:F_kp01} and \ref{figure:F_knp02} respectively.  Already for these values of $\Omega_k$ one can see that the GR point (shown where the two dot-dashed lines cross in each figure) is moving away from the best-fit point.  In fact, in the case where $\Omega_k = -0.02$, for the functional form evolution there is very nearly a deviation from GR at the 95\% level.  This shows that indeed ignoring curvature can possibly produce a false positive for a deviation from general relativity when higher precision data is used.

\begin{figure}[t]
\begin{center}
\begin{tabular}{|c|}
\hline 
{\includegraphics[width=4.33in,height=1.25in,angle=0]{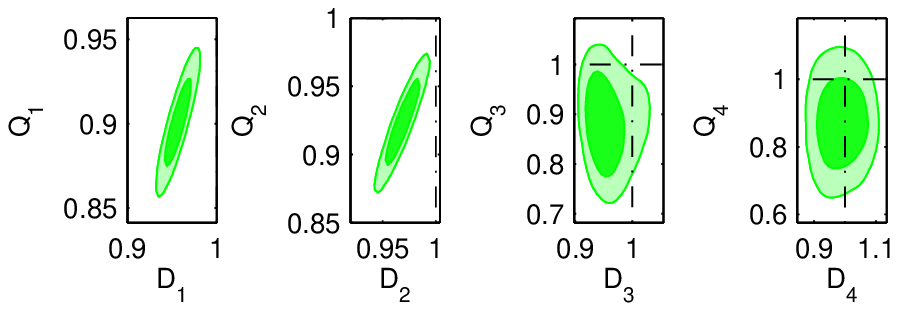}}\\ \hline
{\includegraphics[width=4.33in,height=1.25in,angle=0]{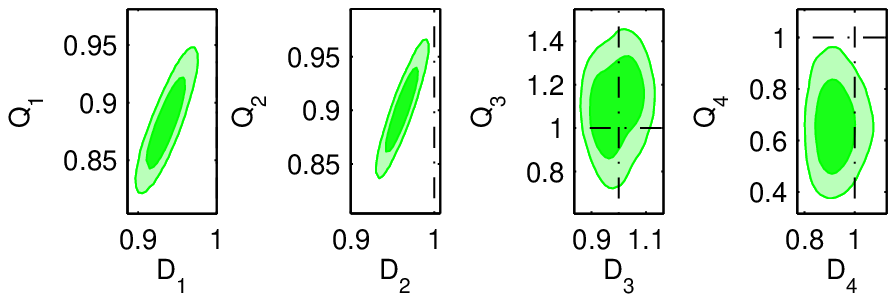}}\\ \hline
\end{tabular}
\\
\begin{tabular}{|c|}
\hline
{\includegraphics[width=2.16in,height=1.25in,angle=0]{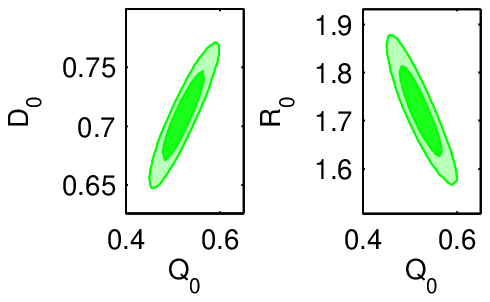}} \\
\hline   
\end{tabular}
\caption{\label{figure:F_kp05}  
Plotted are the 68\% and 95\% 2D confidence contours for the MG parameters when a spatially flat model is used but the actual underlying universe has $\Omega_k=0.05$.
TOP: Confidence contours for the MG parameters $Q_i$ and $\mathcal{D}_i$, $i=1,..4$ when the traditional binning method is used.  All parameter values are pulled to smaller values and indicate a deviation from general relativity.  MIDDLE: Confidence contours for the MG parameters $Q_i$ and $\mathcal{D}_i$, $i=1,..4$ when the hybrid evolution method is used.  Most of the parameter contours are pulled to smaller values and indicate a deviation from general relativity. BOTTOM:  Confidence contours for the MG parameters $Q_0$ and $\mathcal{D}_0$ and $R_0$ when the functional form binning method is used.  A strong deviation (due to the assumption of spatial flatness) from reneral Relativity is present in both contours.}  
\end{center}
\end{figure}

\begin{figure}[t]
\begin{center}
\begin{tabular}{|c|}
\hline 
{\includegraphics[width=4.33in,height=1.25in,angle=0]{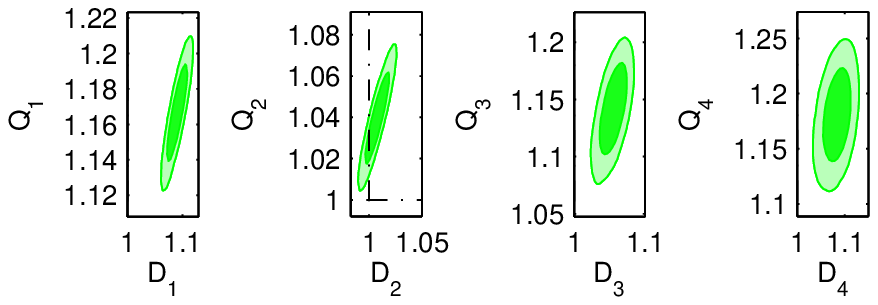}}\\ \hline
{\includegraphics[width=4.33in,height=1.25in,angle=0]{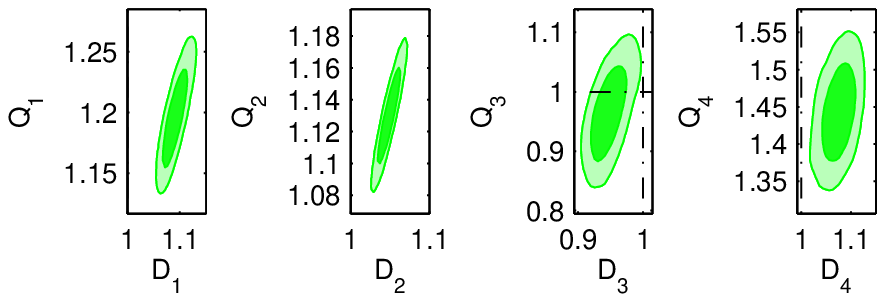}}\\ \hline
\end{tabular}
\\
\begin{tabular}{|c|}
\hline
{\includegraphics[width=2.16in,height=1.25in,angle=0]{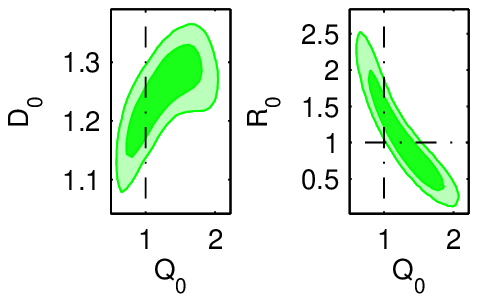}} \\
\hline   
\end{tabular}
\caption{\label{figure:F_knp1} 
Plotted are the 68\% and 95\% 2D confidence contours for the MG parameters when a spatially flat model is used but the actual underlying universe has $\Omega_k=-0.1$
TOP: Confidence contours for the MG parameters $Q_i$ and $\mathcal{D}_i$, $i=1,..4$ when the traditional binning method is used.  All parameter values are pulled to larger values and indicate a deviation from general relativity.  MIDDLE: Confidence contours for the MG parameters $Q_i$ and $\mathcal{D}_i$, $i=1,..4$ when the hybrid evolution method is used.  Most of the parameter contours are pulled to larger values and indicate a deviation from general relativity.  BOTTOM:  Confidence contours for the MG parameters $Q_0$ and $\mathcal{D}_0$ and $R_0$ when the functional form binning method is used.  The $Q_0-\mathcal{D}_0$ shows a deviation from general relativity, while the $Q_0-R_0$ contour shows some tension with the GR point.}   
\end{center}
\end{figure}

\begin{samepage}
To further explore how large an impact assuming the universe is flat when in fact it is curved will have on constraints on the MG parameters four more fiducial data sets were included with $\Omega_k=\pm 0.05$ and $\Omega_k=\pm 0.1$.  Though these values for the curvature parameter $\Omega_k$ are well outside current constraints from all combined data, these constraints on $\Omega_k$ were obtained by using a $\Lambda$CDM model, and while in the $\Lambda$CDM cosmological model observations require a universe that is flat or very close to it, modified theories of gravity or inhomogeneous cosmological models may require the universe to be significantly more curved to fit observations.  Additionally, given the fact that our simulated data sets do not represent the constraining power of future experiments, it is useful to use these larger values of the curvature parameter to illustrate how constraints from data sets with smaller uncertainties or more data will be affected.  While the data sets we simulated and used may require these large curvature values for a significant apparent deviation to arise, future data sets with more precise measurements may not. For most of these values of $\Omega_k$, we find for every parameter evolution method (traditional binning, hybrid method, or functional form) there is at least one MG parameter and in most cases more than one MG parameter that strongly deviates from its GR value.  Almost all $Q-\mathcal{D}$ 2D parameter contours for these values of $\Omega_k$ show the GR, $(1,1)$, point outside their 95\% confidence limits constraints.  In Figs. \ref{figure:F_kp05} and \ref{figure:F_knp1} the 2D confidence contours for the $\Omega_k=0.05$ and $\Omega_k=-0.1$ fiducial data sets are shown respectively.  Interestingly the negatively curved model ($\Omega_k=0.05$) deviates from GR much more substantially than does the closed model for all evolution methods of the MG parameters.  This could be due to the way the comoving angular diameter distance enters into both the weighting factor for the weak lensing as well as the way the wave number, $k$, is determined when calculating the lensing cross power spectrum.  
\end{samepage}
\begin{center}
\begin{table}[h!]
\begin{tabular}{|c|c|c|c|c|c|c|c|}\hline
\multicolumn{8}{|c|}{{ \bfseries Correlation coefficients between $\Omega_k$ and the MG parameters}}\\ \hline
\multicolumn{8}{|c|}{{MG parameters evolved using traditional binning}}\\ \hline
$Q_1$&$Q_2$&$Q_3$&$Q_4$&$\mathcal{D}_1$&$\mathcal{D}_2$& $\mathcal{D}_3$&$\mathcal{D}_4$\\ \hline
\,0.3783 \,& \, 0.1289 \,  & \,  0.1201 \, &  \, 0.0074 \,  &  \, 0.3135 \,  & \,  0.0492 \,  & \,  0.0748  \, & \,  -0.0102 \,  \\ \hline
\multicolumn{8}{|c|}{}\\ \hline
\multicolumn{8}{|c|}{{ MG parameters evolved using hybrid binning}}\\ \hline
$Q_1$&$Q_2$&$Q_3$&$Q_4$&$\mathcal{D}_1$&$\mathcal{D}_2$& $\mathcal{D}_3$&$\mathcal{D}_4$\\ \hline
\,0.4591\, &\,0.1997\, &\,-0.0489\, &\,0.1584\, &\,0.4244\, &\,0.0982\, &\,0.0968\, &\,0.0278\, \\ \hline
\end{tabular}
\\
\begin{tabular}{|c|c|c|c|}\hline 
\multicolumn{3}{|c|}{}\\ \hline
\multicolumn{3}{|c|}{{ MG parameters evolved using the functional form}}\\ \hline
$Q_0$&$\mathcal{D}_0$&$R_0$\\ \hline
\,\,\,\,\,\, \,\,\,0.0289 \,\,\,\,\,\, \,\,\, & \,\,\,\,\,\, \,\,\,0.0969\,\,\,\,\,\, \,\,\,& \,\,\, \,\,\,-0.0095 \,\,\, \,\,\, \\ \hline
\end{tabular} 
\caption{\label{table:FutCorr}
We list the correlation coefficients between $\Omega_k$ and the various MG parameters for a universe with $\Omega_k = 0.05$.  In contrast to those obtained from current observational data, these correlation coefficients are almost all consistent with the trends observed in Fig. \ref{figure:MGPVOK} when a flat universe is assumed but curved simulated future data is used.  }
\end{table}
\end{center}

\begin{figure}[H]
\begin{center}
\begin{tabular}{|c|c|c|}
\hline
Traditional Binning Evolution & Hybrid Evolution & Functional Form Evolution\\
\hline 
{\includegraphics[width=1.9in,height=1.9in,angle=0]{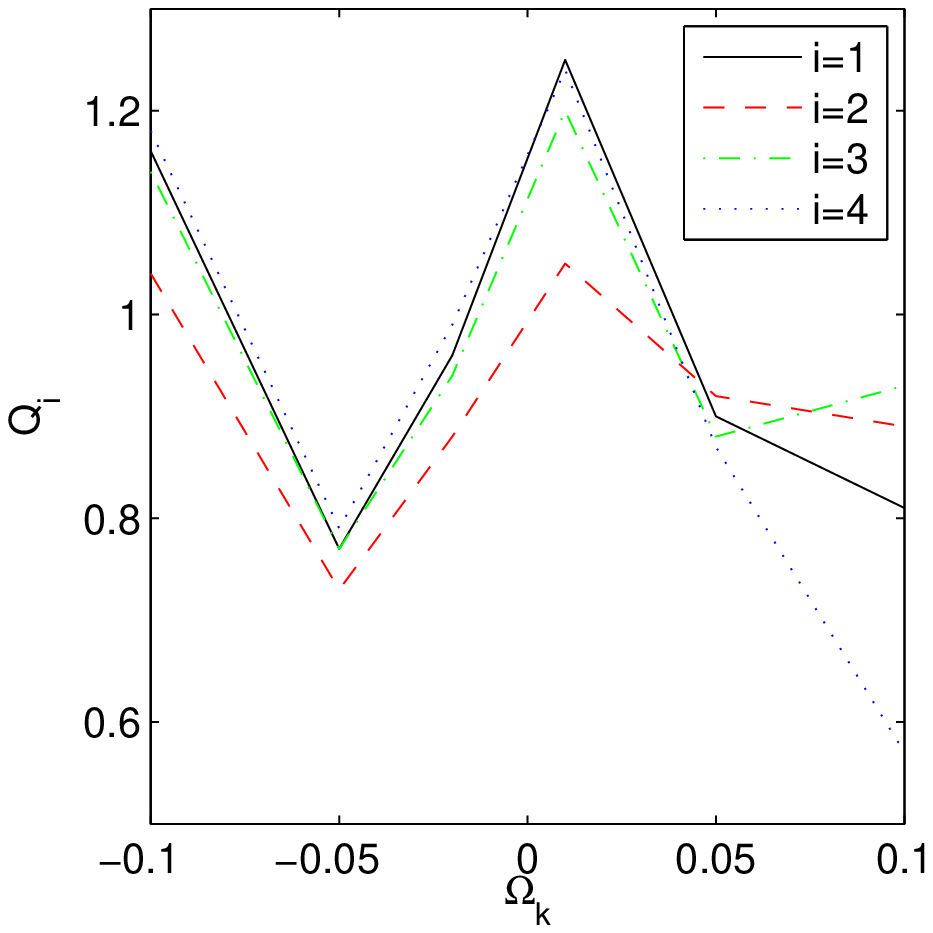}}&
{\includegraphics[width=1.9in,height=1.9in,angle=0]{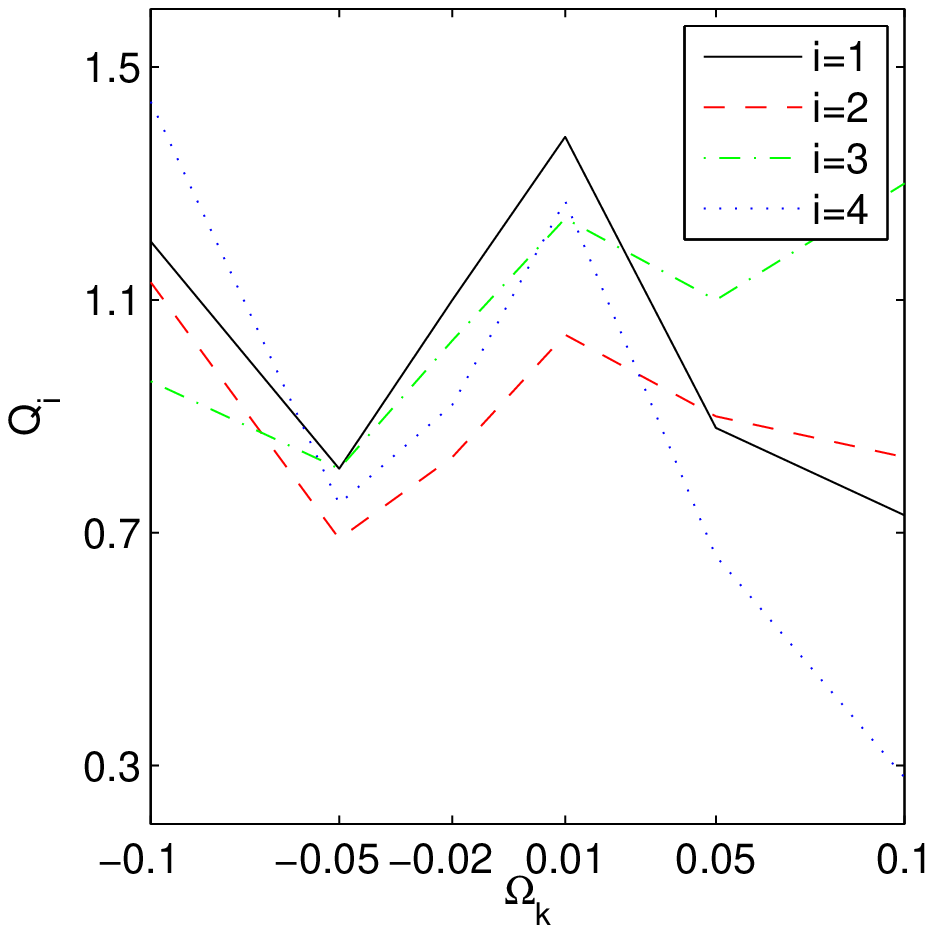}}&
{\includegraphics[width=1.9in,height=1.9in,angle=0]{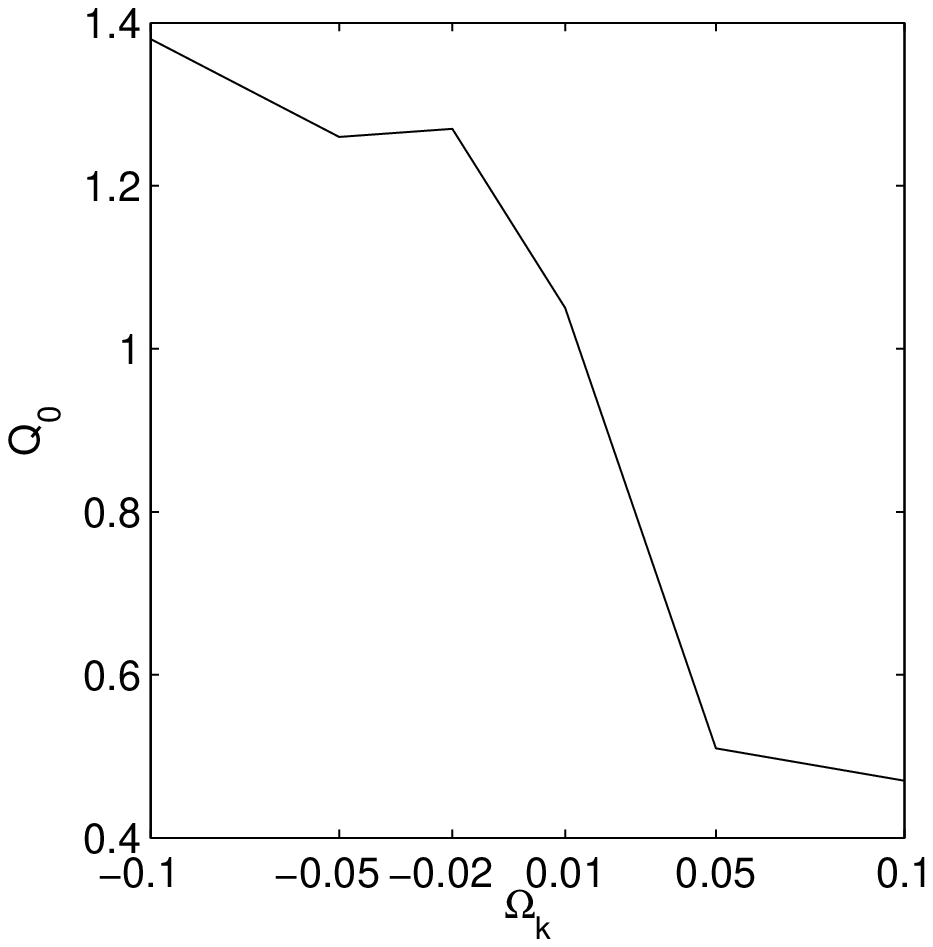}}\\ \hline
{\includegraphics[width=1.9in,height=1.9in,angle=0]{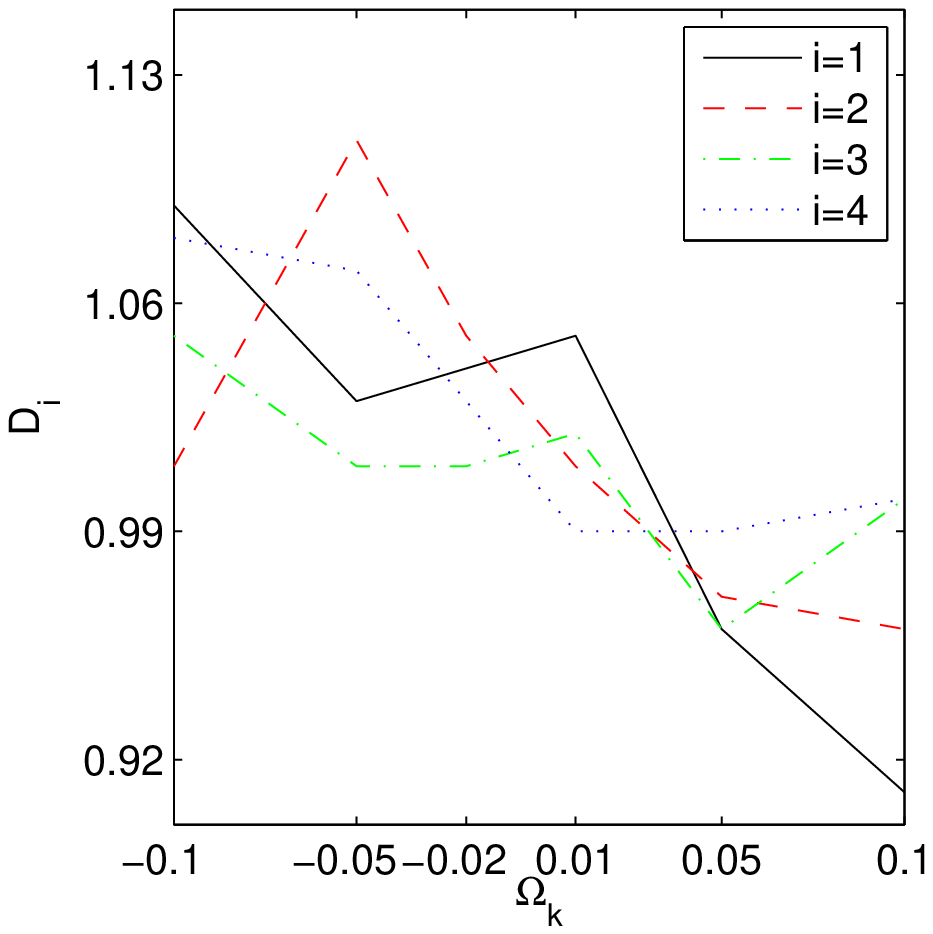}}&
{\includegraphics[width=1.9in,height=1.9in,angle=0]{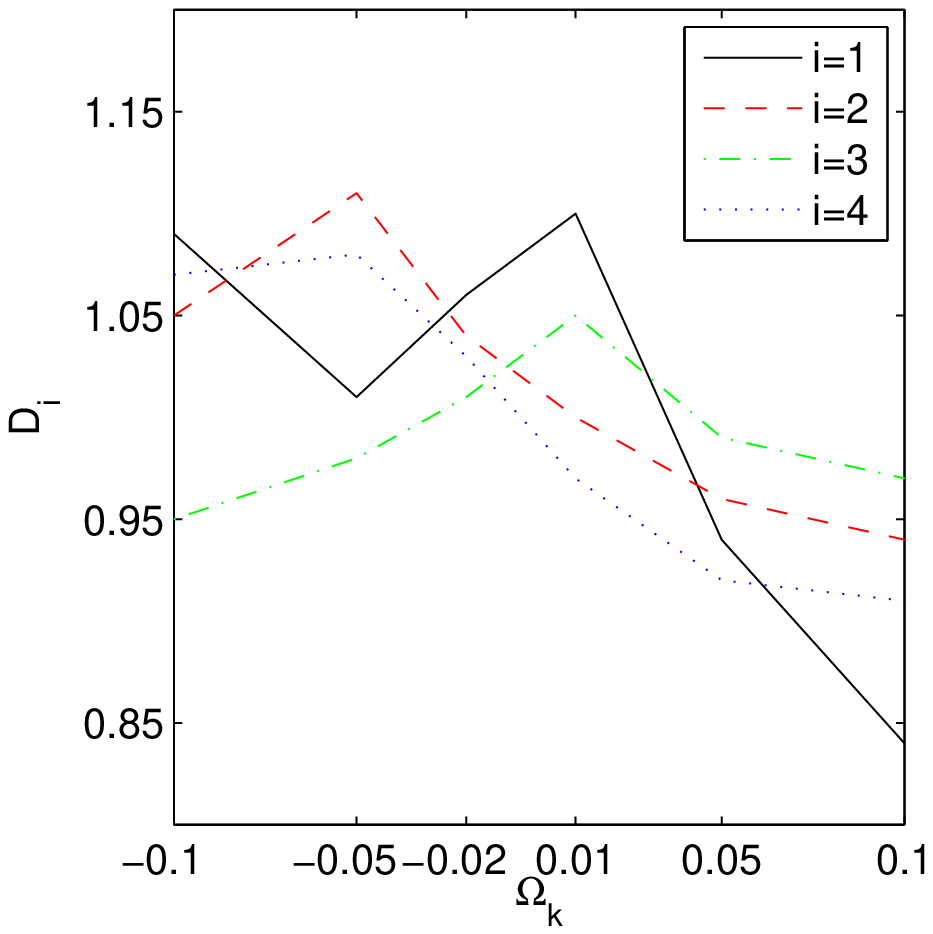}}&
{\includegraphics[width=1.9in,height=1.9in,angle=0]{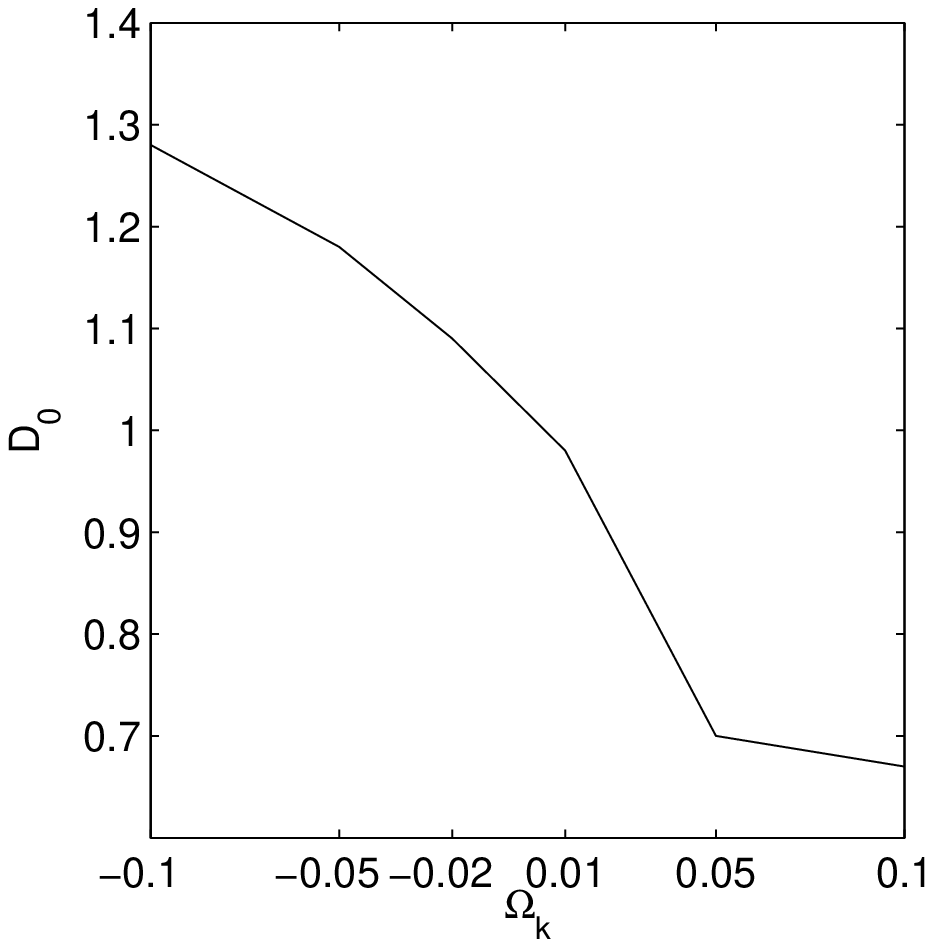}}\\ \hline
\end{tabular}
\caption{\label{figure:MGPVOK} Here we plot the best-fit points for the parameters $Q$ and $\mathcal{D}$ for the various evolution methods versus the underlying $\Omega_k$ fiducial values while a spatially flat universe is assumed.  Most of the trends have a negative slope which is expected from looking at Eqs (\ref{eq:PoissonMod}) and (\ref{eq:PoissonModSum}) (note that this trend is consistent with the positive correlations between the MG parameters and the curvature parameter as explained in Sec. IV-C).  TOP LEFT: Plots for the parameters $Q_i$, $i=1,2,3,4$ for the traditional binning method.  TOP MIDDLE: Plots for the parameters $Q_i$, $i=1,2,3,4$ for the hybrid evolution method.  TOP RIGHT: Plots for the parameters $Q_0$ for the functional form evolution method.  BOTTOM LEFT: Plots for the parameters $\mathcal{D}_i$, $i=1,2,3,4$ for the traditional binning method.  BOTTOM MIDDLE: Plots for the parameters $\mathcal{D}_i$, $i=1,2,3,4$ for the hybrid evolution method.  BOTTOM RIGHT: Plots for the parameters $\mathcal{D}_0$ for the functional form evolution method. } 
\end{center}
\end{figure}

One other feature to notice in all four sets of plots is the directions which the constraint contours move in the parameter space.   Looking at Eqs. (\ref{eq:PoissonMod}) and (\ref{eq:PoissonModSum}), one would expect that for a universe with a positive value of $\Omega_k$ and thus a negative $K$ value the assumption of a flat model would demand smaller values for the parameters $Q$ and $\mathcal{D}$ and vice versa.  This is indeed the trend we see for the most part in Figs. \ref{figure:F_kp01}, \ref{figure:F_knp02}, \ref{figure:F_kp05}, and \ref{figure:F_knp1}.  To further show how fiducial models with higher values of $\Omega_k$ have lower best-fit MG parameters when a flat universe is assumed, in Fig. \ref{figure:MGPVOK} we plot the best-fit MG parameter values versus $\Omega_k$ for the various evolution methods.  As was discussed above most of the parameters exhibit a negatively sloping trend. This seems to suggest that the observed behavior where most of the $Q$ and $\mathcal{D}$ parameters move the same direction in the parameter space could act as a signature of a false positive in the future if flatness is assumed and a deviation from GR is detected.

\begin{samepage}
One can also notice that the trends discussed above do not match the behavior expected from the correlation coefficients obtained when using the current data.  For the $Q$ and $\mathcal{D}$ parameters the behavior described above corresponds to positive correlation because by assuming flatness (for example, in the case of a positive $\Omega_k$) we are forcing a lower value for $\Omega_k$ and thus would expect a lower value for our MG parameters. We do not observe this behavior with the current data.  This is most likely due to the large uncertainties in this data which allows a large number of models to fit quite well, and thus makes the calculation of the correlation coefficients less accurate.  A more accurate description of the correlation coefficients can be obtained by using some that were calculated when running the "null" tests on our fiducial models.  We show the correlation coefficients for the case when $\Omega_k = 0.05$ in Table \ref{table:FutCorr}.  These correlation coefficients are almost all consistent with the trends we see in the constraints as well as the best-fits.  
\end{samepage}

\section{Conclusion}
In this work we extended previous studies and the framework of modified growth (MG) parameters to test general relativity (GR) at cosmological scales in order to include spatial curvature in the models, as while current data when analyzed using the $\Lambda$CDM model points to a universe that is flat or very close to it, this constraint may not hold in modified theories of gravity or inhomogenous cosmological models. Using the latest cosmological data sets we explored the correlations between MG parameters and the curvature parameter $\Omega_k$ finding that indeed there are non-negligible correlations.  We next used future simulated data to explore whether assuming a spatially flat model on a spatially curved universe would affect the MG parameter constraints.  We found that indeed, for our simulated data sets, such an assumption of flatness can cause tension with GR when using an $\Omega_k$ as little as $0.02$ away from the flat case and significant apparent deviations for $\abs{\Omega_k} \ge 0.05$ .  Models with larger departures from flatness cause more MG parameters to deviate from GR with even larger discrepancies.  We also found that negatively curved models deviated more quickly and more significantly from GR when the assumption of flatness is made.  

As expected from the derived modified growth equations, for all approaches and for most of the bins, we find positive correlation coefficients between the MG parameters and $\Omega_k$ from using simulated future data which is much more precise than what is currently available. The trends for the best-fit MG parameters versus fiducial values of $\Omega_k$ used are found to be consistent with these correlations. 
Though the values of the correlation coefficients found are of course dependent upon the parametrization used, the signs of the correlations coefficients for the most part remain consistent across parametrizations.

The results obtained in this analysis show that when using high-precision data from future experiments in order to test general relativity at cosmological scales or look for deviations from it, one must take into account the effect of curvature.  This point may also be relevant when trying to test other, alternative theories of gravity. Indeed, our results indicate that the assumption of a spatially flat universe when performing these tests can bias the MG parameter constraints, leading to apparent deviations from general relativity.

\begin{acknowledgements}
We would like to thank Lucia Popa and Ana Caramete for useful comments about the binning version of the ISiTGR code. M.I. acknowledges that this material is based upon work supported by the Department of Energy (DOE) under grant DE-FG02-10ER41310, NASA under grant NNX09AJ55G, and that part of the calculations for this work have been performed on the Cosmology Computer Cluster funded by the Hoblitzelle Foundation.  J.D. acknowledges that this research was supported in part by the DOE Office of Science Graduate Fellowship Program (SCGF), made possible in part by the American Recovery and Reinvestment Act of 2009, administered by ORISE-ORAU under contract no. DE-AC05-06OR23100.
\end{acknowledgements}

\end{document}